# Decentralized Trajectory Tracking Using Homology and Hodge Decomposition in Sensor Networks

Xiaotian Yin[2], Yu-Yao Lin[1], Chien-Chun Ni[1], Jiaxin Ding[1], Wei Han[2], Dengpan Zhou[1], Jie Gao[1] and Xianfeng Gu[1]

[1]Department of Computer Science, Stony Brook University
[2]Department of Mathematics, Havard University

**Abstract**

With the recent development of localization and tracking systems for both indoor and outdoor settings, we consider the problem of sensing, representing and analyzing human movement trajectories that we expect to gather in the near future. In this paper, we propose to use the topological representation, which records how a target moves around the natural obstacles in the underlying environment. We demonstrate that the topological information can be sufficiently descriptive for many applications and efficient enough for storing, comparing and classifying these natural human trajectories. We pre-process the sensor network with a purely decentralized algorithm such that certain edges are given numerical weights. Then we can perform trajectory classification by simply summing up the edge weights along the trajectory. Our method supports real-time classification of trajectories with minimum communication cost. We test the effectiveness of our approach by showing how to classify randomly generated trajectories in a multi-level arts museum layout as well as how to distinguish real world taxi trajectories in a large city.

## 1 Introduction

Powered by recent technology advancements, real-time target tracking is becoming a reality for both outdoor and indoor scenarios. These technologies adopt various sensing modalities (RF signals, infrared, visible light, etc) and employ a variety of design techniques (device-based v.s. device free, active tracking v.s. passive tracking) [1, 5, 23, 25, 29, 41, 42, 47]. It can be expected that as these tracking systems are gradually in place; we will soon be able to gather a large amount of real world motion data, suggesting an unprecedented opportunity to understand and analyze natural human movement behaviors.

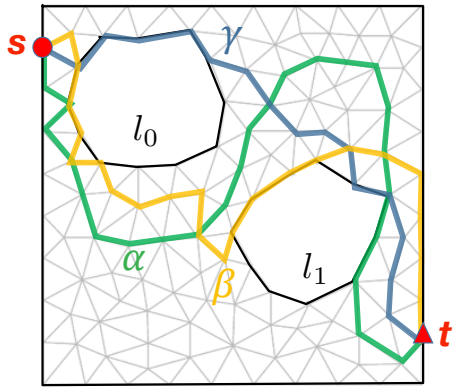

**Figure 1.** The network is with two holes/obstacles marked as black loops ($l_0$, $l_1$). Paths $\alpha$ and $\gamma$, paths $\beta$ and $\gamma$ are different topologically as they get around different subsets of holes, while $\alpha$ and $\beta$ are topologically equivalent.

Consider the following scenario in a museum. Suppose that the museum is instrumented with an indoor tracking system. Each visitor wears a tag that can communicate with the tags broadly installed in each exhibition room. This tag-tag communication allows us to track the detailed movement patterns of the visitors within the museum. Understandings of these motion patterns may lead to the improvement of the display arrangements in the exhibition rooms, ranking the popularity of art pieces, among many others. Take another example of vehicles traveling in a large metropolitan area and tracked by roadside units, these traffic patterns can be of great interest for potential traffic optimization and civil planning. In all cases, new discoveries of human movement patterns are fundamentally interesting and practically useful.

We assume a domain of interest with sensors uniformly deployed in the domain, and locally connected to form a planar graph as in Fig. 1. Notice that this domain could be complicated, it could have big holes or might be in 3D (multi-floor building). In this domain, the mobile entities are detected by the sensors in proximity and such time-stamped detection data are treated as our raw trajectory data. These tracking/monitoring data are a unique type of sensing data: they are sequential (temporal), containing geographical information, typically bulky, and may involve unnecessary details. For example, the exact geometric path from visitors' tag tracking data might be inessential for an exhibit planner who just wants to understand the sequences of room visited; analyzing which lane the taxi drivers stay on might be less important than which intersection they make a turn. In these examples, we actually care more about the topological features of the trajectory (how they move around the obstacles in the domain) rather than the geometrical features of the trajectory (which lane they drive on the street). In fact, most of the time it is the topological features of the trajectory that are of main interest.

To acquire the topological features of the trajectory, one needs to collect,

prune, and process the trajectory data that are inherently collected from the distributed sensors. Generally, the trajectory data is a sequence of the IDs or geographical locations of sensors which a target visited. Existing algorithms do not directly gather the topological information of the trajectories. In one approach, the target detection events and logs are separately stored at individual sensors. While this minimizes the amount of communication between sensors during processing, the global knowledge and the topological features of a trajectory is not present locally. In the second approach, one can possibly construct the trajectory explicitly and propagate this information to the other sensor nodes, facilitating the query for such information. However the data size of a trajectory grows with its length. Thus to pass around a trajectory explicitly, the processing cost, in particular the communication cost, can be prohibitive. In the third approach, events detected on each sensor are delivered to a central server to construct a centralized view. Yet, reporting and retrieving such information from a central server for further process involves a hefty communication cost for the whole network, and the central server represents a single point of failure.

The contribution of this paper is to provide a framework to process and analyze these trajectory data in-network in *real time* with low communication cost, and also support queries, comparisons and classifications of trajectories. One of the crucial ideas is to replace the *geometric* representation of a trajectory, such as GPS logs, by the *topological* representation of a trajectory. There are various definitions of the topological structures of paths. The most commonly used ones are based on homotopy or homology, both count how the 'holes' are enclosed by loops on a surface (or more general topological space). Intuitively homology considers loops as closed curves without orientation, while homotopy treats loops as oriented parametric curves, and in that case it matters how the holes are ordered.

Take the example in Fig. 1. There are two obstacles in the domain and there are several different ways to go from $s$ to $t$. Observe that paths $\alpha$ and $\gamma$, paths $\beta$ and $\gamma$ are both different in a global sense. We cannot deform $\alpha$ to $\gamma$ unless jumping over obstacle $l_0$. However, paths $\alpha$ and $\beta$ are only different in a local manner. One can deform $\alpha$ to $\beta$ smoothly through some local changes. This difference is characterized by the *homotopy type* of a path. Two paths are *homotopy equivalent* if one can be smoothly deformed to the other. In other words, the cycle formed by two paths of the same homotopy type can be shrunk to a point. This homotopy equivalence relation naturally partitions a group of paths into equivalent classes, providing well-defined clusters [4, 30]. In theory, the number of homotopy type is infinite, since one can loop around a hole infinitely many times. But in practice, only a finite number of homotopy types are of interest. All paths of the same homotopy types belong to the same class by homology. Thus homology is a weaker classification of homotopy. The benefit of using homology is that it is much easier to compute in many cases.

In the following we provide a quick overview of our approach and contributions. We survey related work afterwards.

## 1.1 Network Setup

We assume that the sensors are deployed on a surface that might have holes (corresponding to obstacles or other domain features) or high-order topological features such as handles. Let sensors be vertices and the connection between sensors be edges. The sensors form a network. We suppose that the sensors collaboratively apply boundary detection algorithms to identify the nodes adjacent to holes of the domain [10, 11, 14–17, 22, 31, 31, 39, 44], and extract a connected planar subgraph $G$ [18, 20, 38, 46]. This extracted graph $G$ stays on an (unknown) surface $\Sigma$ where the faces corresponding to holes (i.e., boundaries) are locally marked. By these algorithms, each node knows whether it stays on the network boundary or not, but no one has the global knowledge of how many holes there are and where they are. This makes the problem of detecting and comparing the topological properties of two paths, using only local information, to be particularly challenging. One of the main contributions of this paper is to address this problem.

## 1.2 Harmonic Forms

Consider a graph $G$ with a planar embedding on a surface $\Sigma$. The discrete *differential 1-form* [33] is a function $\omega$ defined on directed edges. The value $\omega(a,b)$ for an edge $ab$ is the negation of the value $\omega(b,a)$ for edge $ba$. Now we consider the dual graph $\tilde{G}$. Each face of the graph $G$ corresponds to a node in the dual graph. An edge is placed on two nodes in the dual graph if and only if the two corresponding faces in the primal graph share one edge. A differential 1-form $\xi$ on the graph $G$ can be extended to the dual $\tilde{G}$. The value on an edge in the dual graph is the value of the corresponding edge in the primal graph. A differential 1-form is called a *harmonic 1-form* if it satisfies two properties:

1. it is *divergence-free*: $\forall u$ of $G$, $\sum_{v \in N(u)} \omega(u,v) = 0$, where $N(u)$ is the set of neighbors of $u$ in $G$;
2. it is *curl-free*, that is, for any node $\tilde{u}$ of the dual graph $\tilde{G}$, $\sum_{\tilde{v} \in N(\tilde{u})} \omega(\tilde{u}, \tilde{v}) = 0$, where $N(\tilde{u})$ is the set of neighbors of $\tilde{u}$ in $\tilde{G}$.

The first property means that a harmonic 1-form does not have any sources or sinks. If we consider a harmonic 1-form as a flow vector defined on each edge, we have the flow conservation property at each node – what flows in equals what flows out. The second property means that the integration of a harmonic 1-form along any face of $G$ is zero, i.e., in the dual graph there are no sources or sinks either.

With the help of a harmonic 1-form, we can easily test whether two paths are homologous. In particular, we connect the two paths $\alpha, \beta$ with same starting and end positions as a cycle $\alpha - \beta$ (with $\beta$ in reversed direction). If the two paths are homologous, the cycle encloses no holes/handles. By the definition of the harmonic 1-form, if we sum up the weights of the edges along the cycle in clockwise order, the summation must be zero. This represents an extremely simple homology test for the cycle, by only the knowledge of the harmonic 1-form, which can be locally stored on the edges of the network. The communication cost is proportional to the total length of the paths. This is the minimum possible as an algorithm must at least read in the input. Computing the harmonic forms for homology testing is done in a preprocessing phase, by *Hodge decomposition*, which will be explained later. There are infinitely many harmonic 1-forms but only $k$ of them are linearly independent, where $k$ depends on the number of obstacles in the domain (and the number of handles if the domain is not in 2D). These linearly independent ones form a harmonic 1-form basis $\Omega$ on $G$.

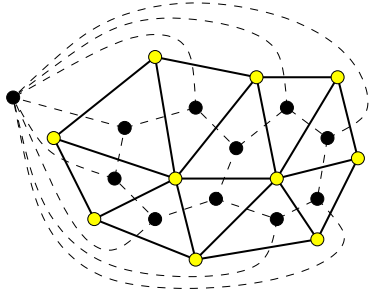

**Figure 2.** A planar graph $G$ and its dual graph $\tilde{G}$. Each face of the graph $G$ corresponds to a node (shown as dark circles) in the dual graph.

We remark that the differential 1-form has been practiced for in-network storage of target tracking data to answer real-time range queries [37]. In that work the differential 1-form is defined by the target presence such that the integration along the edges of a face gives the number of targets inside the face. It is clearly not harmonic (not curl-free). In our case, the harmonic 1-form is independent of target trajectories. Thus we can compute the harmonic 1-forms at the initialization of the network and use them for the entire lifetime of trajectory sensing and queries.

## 1.3 Hodge Decomposition

To compute a harmonic 1-form, we first create an arbitrary 1-form $\omega$, say by randomly assigning weights on the edges of the graph $G$. This 1-form is by no means harmonic. The theory of Hodge decomposition says that for any 1-form, we can decompose it into three components:

$$\omega = df + \delta g + h,$$

where

1. $h$ is a harmonic 1-form, it is divergence-free and curl-free;
2. $df$ is a gradient flow, i.e., there is a potential function $f$ defined on the nodes of $G$ such that $df = f(u) - f(v)$. A gradient flow is curl-free; and

3. $\delta g$ is a curl flow, i.e., the gradient flow in the dual graph. There is a potential function $g$ defined on faces such that $\delta g = g(x) - g(y)$, where $x$ and $y$ are the faces to the right and left of edge $uv$ respectively.

The Hodge decomposition basically states that if we take out the gradient flow and curl flow that contribute to having sources/sinks and curls in the flow, we are left with a harmonic 1-form. In this paper, we develop a purely decentralized algorithm that runs in iterative, gossip style operations that solve the gradient flow $df$ and curl flow $\delta g$. After we subtract them from the 1-form $\omega$, we can obtain the harmonic 1-form $h$. For more details on Hodge decomposition or combinatorial Hodge theory, we refer the reader for the following reference [27, 40].

To acquire the harmonic 1-form basis, we apply the *Hodge decomposition* on the network multiple times with different random initial values on each edge, and testing if the harmonic forms are linearly dependent, which is a simple operation that each sensor can individually test at its own neighborhood. Then we use the first $k$ linearly independent harmonic 1-forms as a harmonic 1-form basis $\Omega = \{\omega_1, \omega_2, \cdots, \omega_k\}$.

## 1.4   Trajectory Sensing and Classification

For target trajectories we first classify them by their beginning and ending positions. In indoor settings the beginning and ending are typically limited by a small number of entrances. For all trajectories that share the same beginning and ending positions we classify them by homology into buckets. For two trajectories in the same bucket, if we glue them as a cycle the loop has trivial homology. In other words, we call trajectories in the same bucket in the same *Trajectory Class (T-class)*. The T-class of a trajectory can be measured by a *T-tuple*: $(s, t, h)$, where $s, t$ are the beginning and ending positions and $h$ is a vector that encodes the information necessary to infer path homology. Here the dimension of $h$ is $k$, the dimension of a harmonic 1-form basis. Here we formally define our trajectory classification: For any two paths $\gamma_1, \gamma_2$ in $G$, we say these paths are in the same *T-class* if the following two conditions are satisfied:

1. $\gamma_1$ and $\gamma_2$ share the same source point $s$ and target point $t$;

2. for any $\omega_i \in \Omega$, $\langle \omega_i, \gamma_1 - \gamma_2 \rangle = \sum_{e_j \in \gamma_1 - \gamma_2} \omega_i(e_j) \leq \mu$. In other words, the summation of harmonic 1-forms along $\gamma_1 - \gamma_2$ is zero or smaller than an error threshold $\mu$.

Since the problem boils down to summing up the values of the harmonic one-forms, we classify the trajectories by homology not homotopy.

We pre-process the network such that identifying the T-tuple can be done in real time by simply summing up the weights of the edges crossed by the trajectory, where the weights are the harmonic 1-forms defined by our pre-processing algorithm. This allows us to quickly perform clustering and trajectory comparison in real time.

We note that our method is purely combinatorial – no geometric embedding is needed beyond having the network represented by a combinatorial planarized domain. This nice property is shared by the algorithm by Ghrist [21] which uses the winding number and angle measurements to test whether a point is inside a cycle. Our method is also purely decentralized. All the network nodes run the same code and none of them does anything special.

In the following, we first briefly review the main technique developed in this paper as well as previous work. We then report our algorithm and evaluation on real data sets. We present examples of visitors in an art museum and real world taxi trajectories that show how trajectories are classified into buckets of different types.

## 2  Related Work

The problem of testing whether two geometric paths $\gamma_1$ and $\gamma_2$ (sharing the same source and target positions on a 2D domain) are homotopic has been studied in the centralized setting. Suppose the obstacles are represented by polygons with a total of $n$ vertices and the paths are given as polygonal curves. A $\Theta(n \log n)$ running time algorithm is available for simple paths and an $O(n^{3/2} \log n)$ time algorithm is known for self-intersecting paths [6]. For paths defined on a general surface, the homotopy test boils down to checking whether the cycle connected by the given two paths is contractible (i.e., shrink to a point). This problem can be solved in a centralized setting in linear running time [12] when the surface and the paths are both available.

In the literature of GIS and data mining, various metrics have been developed to measure the similarity of two trajectories [19, 32] such as dynamic time warping [2], similarity based on longest common subsequence [43], edit distance on real sequence [7], and edit distance with real penalty [8] etc. However, trajectory classification and clustering methods with homotopy/homology are less explored. In [34, 35], a problem defined on configuration space similar to our problem is studied. In their works, the filtration of simplicial complexes defined by persistent homology is used to sample on the nodes. Thereafter, a centralized algorithm is applied to obtain the homology classes of trajectories. Other persistent homology based trajectory classification method can be found in [3, 28]. In [9], the harmonic form generated by Laplacian is applied to detect homology cycle for hole coverage. While this method also applied harmonic form, it requires Laplacian matrix that need to be centralized updated when the network is divided into two sub networks. More detailed classification and clustering methods are reviewed in [48]. Other methods include dynamic time warping [2], similarity based on longest common subsequence [43], edit distance on real sequence [7], and edit distance with real penalty [8] etc.

For a distributed solution for trajectory classification, there are a number of possible schemes when the domain is in 2D. For example, in a 2D setting we can connect each obstacle $i$ by a path $\lambda_i$ to the outer boundary. Then for each trajectory we record the sequence of intersections with $\lambda_i$, together with the direction of crossing. However, this scheme only works for 2D domains and does not apply to general cases. Further, the storage required for this representation is proportional to the number of times that the target trajectory intersects these paths $\lambda_i$, which depends on how $\lambda_i$ are selected and can be suboptimal. In comparison, the approach of harmonic one-forms applies to general domains.

## 3 Preliminaries

In this section, we formally define our domain structure in a combinatorial manner. In the introduction we defined differential 1-forms on planar graphs (with certain faces marked as holes). Formally we will work with the general setting of an abstract simplicial complex.

### 3.1 Abstract Simplicial Complex

Abstract simplicial complex [13, 36] is described in terms of sets. In our setting, we represent the network connectivity graph $G$ by an abstract simplicial complex. Then we define homology group and cohomology group of a simplicial complex.

**Definition 3.1 (Abstract Simplicial Complex).** *An abstract simplicial complex $K$ is a collection of non-empty subsets of a given node set $V$, such that*

1. *if $v \in V$, then $\{v\} \in K$;*
2. *if $\sigma \in K$ and $\sigma' \subset \sigma$, then $\sigma' \in K$.*

A *simplex* $\sigma \in K$ having $k+1$ distinct nodes $\{v_0, \cdots, v_k\}$ is called a $k$-simplex. The dimension of $\sigma$ is $\dim \sigma = k$. The dimension of $K$ is the maximum dimension of any of its simplices. A graph $G = (V, E)$ can be regarded as an abstract simplicial complex which consists of all the cliques in $G$. In this paper, we adopt 2-dimensional simplicial complex to represent our network graph $G$.

### 3.2 Simplicial Homology

For a simplicial complex $K$, a $k$-chain is a linear combination of $k$-simplices of $K$. All $k$-chains form a linear space, the so-called chain space

$$C_k := \{\sum_{\sigma \in K} \lambda(\sigma)\sigma\},$$

where $\lambda(\sigma) \in \mathbb{R}$. The boundary operator $\partial_k : C_k \to C_{k-1}$ is a linear operator such that

$$\partial_k(\sum_i \lambda_i \sigma_i) = \sum_i \lambda_i \partial_k \sigma_i, \lambda_i \in \mathbb{Z}.$$

and

$$\partial_k[v_0, \cdots, v_k] := \sum_i (-1)^i [v_0, \cdots, v_{i-1}, v_{i+1}, \cdots, v_k].$$

If we consider a 1-simplex $[v_i, v_j]$ and a 2-simplex $[v_i, v_j, v_k]$, then the boundary operator $\partial$ takes the boundary of a simplex:

$$\begin{aligned} \partial[v_i, v_j] &= v_j - v_i, \\ \partial[v_i, v_j, v_k] &= [v_j, v_k] - [v_i, v_k] + [v_i, v_j]. \end{aligned}$$

Chains with empty boundary are called *closed chains*. For example, closed 1-chains are closed cycles in a graph. The chains which are the boundary of other chains are called *exact chains*. For example, an exact 1-chain is a boundary chain of a 2-chain. It can be shown that exact chains are closed, namely $\partial_{k-1} \circ \partial_k = 0$, but closed chains may not be exact [24, 45].

**Definition 3.2 (First Homology Group).** *The first homology group of a simplicial complex $K$ is defined as the quotient group,*

$$H_1(K, \mathbb{Z}) := \frac{\operatorname{Ker} \partial_1}{\operatorname{Img} \partial_2}.$$

Two closed cycles are said to be *homologous* if they enclose the same obstacles or holes. Intuitively, the first homology group characterizes those closed cycles which are not boundaries of 2-chains. Hence, this captures the topological structures of all dimensions. For a planar graph with $k$ holes, the first homology group is of dimension $k$.

### 3.3 Cohomology

The co-chain space $C^k$ is the *dual* space of the chain space $C_k$,

$$C^k := \{\omega : C_k \to \mathbb{R} \mid \omega \text{ is linear}\}.$$

The elements in $C^k$ are called $k$-forms.

The co-boundary operator $d_k : C^k \to C^{k+1}$ is defined as follows: suppose $\omega \in C^k$ is a $k$-form, $\sigma \in C_{k+1}$ is a $k+1$-chain, then $d_k\omega \in C^{k+1}$ is a $k+1$-form,

$$d_k\omega(\sigma) := \omega(\partial_{k+1}\sigma).$$

As an example, the co-boundary operator $d$ operates on a 0-form $f$, where $[v_i, v_j]$ is a 1-simplex. Then $df$ is a 1-form,

$$\begin{aligned} df([v_i, v_j]) &= f(\partial[v_i, v_j]) \\ &= f(v_j - v_i) \\ &= f(v_j) - f(v_i). \end{aligned}$$

If $d_k\omega$ is zero, then $\omega$ is called a *closed form*. If a $k$-form $\omega = d\tau$ for some $k-1$-form $\tau$, then $\omega$ is called an *exact form*. Similarly to homology, exact forms are closed, $d_{k+1} \circ d_k = 0$ [24].

**Definition 3.3 (First Cohomology Group).** *The first cohomology group of a simplicial complex $K$ is defined as the quotient group,*

$$H^1(K, \mathbb{R}) := \frac{Ker\, d_1}{Img\, d_0}.$$

The summation of a closed 1-form $\omega$ around the boundary of a 2-chain is zero,

$$\begin{aligned} d\omega([v_i, v_j, v_k]) &= \omega(\partial[v_i, v_j, v_k]) \\ &= \omega([v_j, v_k] - [v_i, v_k] + [v_i, v_j]) \\ &= \omega([v_j, v_k]) - \omega([v_i, v_k]) + \omega([v_i, v_j]) \\ &= 0. \end{aligned}$$

The co-differential operator $\delta$ is defined as follows. Let $\omega$ be a 1-form, then $\delta\omega$ is a 0-form,

$$\delta\omega(v_i) := \sum_{[v_i, v_j] \in E} \omega([v_i, v_j]).$$

Let $g$ be a 2-form, then $\delta g$ is a 1-form,

$$\delta g([v_i, v_j]) := g([v_i, v_j, v_k]) - g([v_j, v_i, v_l])$$

, where $[v_i, v_j, v_k]$ and $[v_j, v_i, v_l]$ share the 1-chain $[v_i, v_j]$. The forms $\delta\omega$ and $\delta g$ are called *co-exact* form.

Suppose $\omega$ is a 1-form and $\gamma = \{e_i\}$ is a path on $G$. We evaluate the summation of $\omega$ along $\gamma$,

$$\langle \omega, \gamma \rangle := \sum_{e_i \in \gamma} \omega(e_i).$$

In fact, for a path $\gamma$ with source $s$ and target $t$, this is how we assign values to the $h$ of the T-tuple $(s, t, h(\gamma))$. The $j$th component of the vector $h$ is evaluated by $\langle \omega_j, \gamma \rangle$.

### 3.4 Hodge Decomposition

**Definition 3.4 (Harmonic 1-form).** *Suppose $\omega$ is a 1-form, if both $d\omega = 0$ and $\delta\omega = 0$, then $\omega$ is called a harmonic 1-form.*

According to the Hodge theory [26], a 1-form $\omega$ can be uniquely decomposed to

$$\omega = df + \delta g + h, \tag{1}$$

where $f$ is a 0-form, $g$ is a 2-form, $h$ is a harmonic 1-form. Note that each cohomological class has a unique harmonic 1-form. Moreover, the homology and cohomology groups are not only dual notions, but they are also isomorphic; so the dimension of their basis are equal. Finally, we therefore conclude that the harmonic 1-form basis of a planar graph with $k$ holes is $k$.

# 4 Distributed Algorithms

In this section we present distributed implementation for computing harmonic 1-forms and for trajectory classification.

## 4.1 Double Covering

Technically, the following algorithm is operated on the graphs without boundaries. With the help of *double covering*, the algorithm is also adaptable to a graph with boundaries.

The double covering of a graph with boundaries is constructed as follows: 1) make a copy of the graph; 2) cohere the graphs together along their boundaries. In this way, graphs with or without boundaries operate the same under our algorithms.

## 4.2 Hodge Decomposition

We propose a decentralized Hodge decomposition algorithm to compute a harmonic 1-form from a randomly generated 1-form. To generate an input 1-form $\omega$, each node $v_i$ of the network is endowed with a random number $u_i$ in the range of $[-1.0, 1.0]$; then a 1-form $\omega$ is defined over each edge $[v_i, v_j]$, $(i < j)$ by $(u_i + u_j)/2$ in the direction from $v_i$ to $v_j$, and the inverse in the opposite direction.

According to the Hodge theory, $\omega = d\mathbf{f} + \delta\mathbf{g} + \mathbf{h}$. In order to compute the harmonic 1-form $\mathbf{h}$, we need to compute the zero-form $\mathbf{f}$ and two-form $\mathbf{g}$, so that we can further compute the exact component $d\mathbf{f}$, co-exact component $\delta\mathbf{g}$ and harmonic component $h = \omega - df - \delta g$ accordingly. Recall that $d \circ d = 0$, $\delta \circ \delta = 0$. Applying $d$ and $\delta$ on both sides of the equation separately, we have

$$\delta d\mathbf{f} = \delta\omega \qquad (2)$$
$$d\delta\mathbf{g} = d\omega \qquad (3)$$

Equation (2) gives the solution to the zero-form $\mathbf{f}$. By the definition of the operators $d$ and $\delta$, this is equivalent to a linear equation

$$\sum_{v_j \in N(v_i)} (\mathbf{f}(v_j) - \mathbf{f}(v_i)) = \sum_{v_j \in N(v_i)} \omega([v_i, v_j]), \forall v_i \in G,$$

where $N(v_i)$ is the set of neighboring nodes of $v_i$. This system can be solved by sensor nodes in parallel using an iterative method: initialize $\mathbf{f}(v_i) = 0$ for all $v_i$, then keep updating $\mathbf{f}(v_i)$ as follows

$$\mathbf{f}(v_i) \leftarrow \frac{\sum_{v_j \in N(v_i)} \mathbf{f}(v_j) - \sum_{v_j \in N(v_i)} \omega([v_i, v_j])}{|N(v_i)|}, \qquad (4)$$

where $|N(v_i)|$ is the size of $N(v_i)$ (i.e. degree of $v_i$ in $G$). After iterations, $\mathbf{f}$ will converge to the solution. In practice, we stop the iterations at node $v_i$ when the difference of $\mathbf{f}(v_i)$ drops below a user specified threshold $\varepsilon$.

Similarly, Equation (3) gives the solution to the two-form **g**. In the sensor network setting, it is equivalent to a linear system that can be solved iteratively at each triangular face $f_i \in G$. We first initialize $\mathbf{g}(f_i) = 0$ for all $f_i$, then keep updating $\mathbf{g}(f_i)$ as follows until the change of $\mathbf{g}(f_i)$ drops below threshold $\varepsilon$:

$$\mathbf{g}(f_i) \leftarrow \frac{\sum\limits_{f_j \in N(f_i)} \mathbf{g}(f_j) - \sum\limits_{e_k \in \partial f_i} \omega(e_k)}{|N(f_i)|}. \tag{5}$$

In here $N(f_i)$ is the set of faces sharing a common edge with $f_i$, and $\partial f_i$ is the set of edges $\{e_k\}$ adjacent to (counterclockwise) $f_i$.

In particular, if the input graph $G$ is with boundaries, then **f** and **g** are computed on its double covering $\tilde{G}$. Due to the symmetry of $\tilde{G}$, both **f** and **g** are symmetric along the loops which are boundaries of $G$.

Once we solved **f** and **g**, the exact component $d\mathbf{f}$ can be computed on every edge $e_k \in M$ as $d\mathbf{f}(e_k) = \mathbf{f}(v_j) - \mathbf{f}(v_i)$, where $v_i$, $v_j$ are the starting and ending nodes of $e_k$. The co-exact component $\delta\mathbf{g}$ can be acquired by $\delta\mathbf{g}(e_k) = \mathbf{g}(f_i) - \mathbf{g}(f_j)$, where $f_i$, $f_j$ are the faces on the left and right side of $e_k$. Then the harmonic component is simply

$$\mathbf{h}(e_k) = \omega(e_k) - d\mathbf{f}(e_k) - \delta\mathbf{g}(e_k), \forall e_k \in G.$$

Note that errors in **h** depend on the threshold $\varepsilon$ since we apply a numerical algorithm. Also, the number of iterations also depends on the parameter $\varepsilon$.

Our algorithm operates as an iterative algorithm using a gossip style. In particular, assume all nodes are synchronized with slotted time steps. In each round, each node sends its current value to all neighboring nodes. Each node computes a new value. As can be seen in Equation (4) and (5), the formulas to compute the zero-form **f** and two-form **g** are standard gossip algorithms for computing averages, in which **f** is computed on the primal graph $G$ and **g** is computed on the dual graph $\tilde{G}$.

### 4.3 Harmonic 1-Form Basis

For the trajectory classification on a graph $G$ with $k = m$ holes, we need a harmonic 1-form basis $\Omega = \{\omega_1, \cdots, \omega_m\}$ that spans the entire linear space of harmonic 1-forms on $G$. Notice that the value $m$ depends only on the number of holes of $G$, it is a global parameter but in our algorithm sensors can work without such knowledge.

To compute a harmonic 1-form basis $\Omega$, we iteratively call the Hodge decomposition algorithm in Sec. 4.2. Since each 1-form is randomly initialized, the $k$ generated harmonic 1-forms $\Omega' = \{\omega_1, \cdots, \omega_k\}$ are linearly independent with high probability when $k \leq m$. We can therefore acquire a harmonic 1-form basis $\Omega$ once a sufficient amount of generated 1-forms are verified to be spanned by $\Omega'$. Namely, we have $\Omega = \Omega'$.

To confirm whether a harmonic 1-form is linearly independent of $\Omega'$, we check the linear dependency as follows. At each node $v$ in the network, we pick $m'$ edges $\{e_1, e_2, \cdots, e_{m'}\}$ from its local neighborhoods, where $m' \geq \dim H_1(M, \mathbb{Z})$. Notice that the edges further than 1-hub away can also be selected if $degree(v) < m'$. For each harmonic 1-form $\omega_i$, the node constructs a column vector

$$w_i := (\omega_i(e_1), \omega_i(e_2), \cdots, \omega_i(e_{m'}))^T.$$

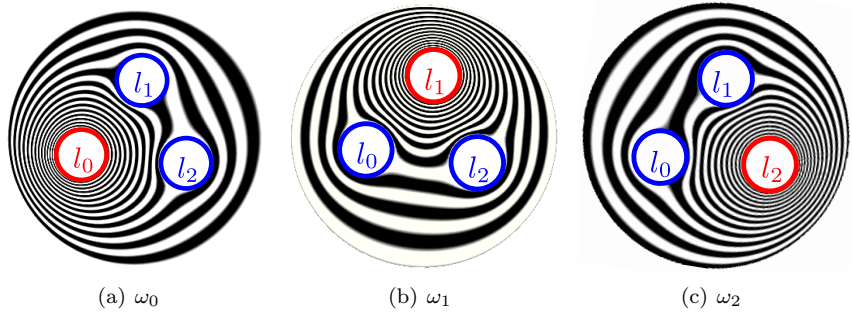

**Figure 3.** Basis for the canonical harmonic 1-forms. Each $\omega_i$ is dual to the interior boundary $l_i$.

At step $k$, the node computes the rank of $\{w_1, w_2, \cdots, w_k\}$; if the rank equals $k$, these 1-forms are linearly independent; Otherwise, they are dependent. The number of holes is given by

$$\min_k \mathrm{Rank}\{w_1, w_2, \cdots, w_k\} = \mathrm{Rank}\{w_1, w_2, \cdots, w_k, w_{k+1}\}.$$

Fig. 3 demonstrates 3 vectors of harmonic 1-form basis for a 3-hole figure domain. The three generators of this basis are actually canonical. That is to say, the harmonic 1-forms $\omega_0$, $\omega_1$, $\omega_2$ are dual to the boundaries of the holes $l_0$, $l_1$, $l_2$ respectively. For any closed loop $l'_i$ homologous to $l_i$, $\langle \omega_i, l'_i \rangle$ is nonzero; while for any loop $l'_j$ that does not enclose $l_i$, $\langle \omega_i, l'_j \rangle = 0$.

### 4.4 Trajectory Classification

Once we have completed harmonic 1-form basis $\Omega = \{\omega_1, \cdots, \omega_m\}$, we can check if two paths are with the same *T-tuple*. Given two paths $\gamma_1$, $\gamma_2$ sharing the same pair of source and destination, we can compute the sum of $\omega_k$ along loop $\gamma_1 - \gamma_2$ as follows:

$$\langle \omega_k, \gamma_1 - \gamma_2 \rangle := \sum_{e_i \in \gamma_1} \omega_k(e_i) - \sum_{e_j \in \gamma_2} \omega_k(e_j)$$

By using harmonic 1-forms, we can detect if a cycle is trivial, i.e., homotopic to a point. We start with the following theorem in a continuous manner.

**Theorem 4.1.** *Suppose a harmonic 1-form basis $\{\omega_k\}_{k=1}^n$ have been obtained by the above algorithm. Let $\gamma$ be a closed chain on the network, then $\gamma$ is homotopic to a point, if and only if $\int_\gamma \omega_k = 0$, for all $k$.*

**Proof:** Suppose the interior boundary components are $\{\gamma_1, \gamma_2 \cdots, \gamma_n\}$, which form a basis of the homology group $H_1(M, \mathbb{Z})$. The dual cohomology basis are $\{\eta_1, \eta_2, \cdots, \eta_n\}$, such that

$$\int_{\gamma_i} \eta_j = \delta_i^j, \tag{6}$$

where $\delta_i^j$ is the Kronecker symbol. According to Hodge theory, each cohomological class has a unique harmonic 1-form, we can assume $\eta_k$'s are harmonic. Because $\{\omega_k\}_{k=1}^n$ is a basis, so

$$\eta_i = \sum_j \lambda_{ij} \omega_j, \omega_i = \sum_j \lambda^{ij} \eta_j, \tag{7}$$

where $(\lambda^{ij})$ is the inverse of $(\lambda_{ij})$. Assume $\gamma = \sum \alpha_i \gamma_i$, $\gamma$ is homotopic to zero, if and only if $\alpha_i = 0, \forall i$. By

$$\alpha_i = \int_\gamma \eta_i \tag{8}$$

and

$$\int_\gamma \eta_i = \sum_j \lambda_{ij} \int_\gamma \omega_j, \forall i,$$

$(\lambda_{ij})$ is invertible, we get $\int_\gamma \omega_j = 0, \forall j$.

□

Ideally, the given two paths are with the same T-tuple if and only if the integration along $\gamma_1 - \gamma_2$ is zero for every $\omega_k \in \Omega$. But our iterative calculation of harmonic 1-forms may carry numerical errors. To tolerate these errors, we use a threshold $\mu$. $\gamma_1$ and $\gamma_2$ are classified to be with the same T-tuple if and only if

$$|\langle \omega_k, \gamma_1 - \gamma_2 \rangle| < \mu, \forall \omega_k \in \Omega.$$

This threshold $\mu$ should be specified as an input to the algorithm. In section 5 we will analyze the safe value of $\mu$ and how the algorithm correctness is affected by $\mu$ and $\varepsilon$.

### 4.5 Trajectory Representation by T-tuple

With the harmonic basis, we can represent each trajectory $P$ by T-tuple $(s, t, h)$, where $h$ is a vector of dimension $m$ representing the summation of the harmonic 1-form $\omega_i$ along this trajectory, for each basis in $\Omega$. Since we expect that the number of holes to be typically constant, the trajectory can be represented in a compact way whose size depends only on the topological complexity of the domain and is independent of the geometric resolution. In addition, natural human movement trajectories have patterns. It is expected that not all different T-class actually show up. For example, a trajectory that loops around an obstacle infinitely many times will not show up in reality. Thus if there are only $r$ different T-classes in real data sets, we expect only $r$ different values for each pair of source and destination. Using a simple hash function we can represent the $r$ different values using space $O(\log r)$. From this perspective, our storage need for representing all the different trajectories, up to the homology accuracy, matches the lower bound.

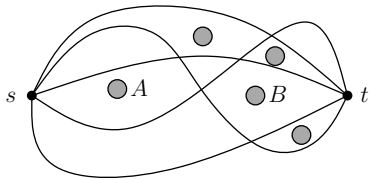

**Figure 4.** If we aim for better descriptive capabilities, we may artificially add all shaded obstacles as holes of the network. Or, if the applications ask for classifications into coarse groups, we may select only two of the obstacles $A$, $B$ such that the trajectories fall into only 4 different categories.

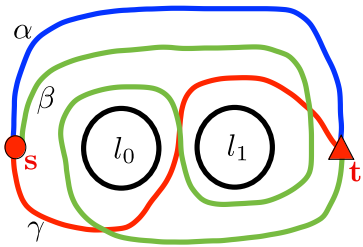

**Figure 5.** An example of trajectories pairs with the same homology but with different homotopy.

In certain applications, users can possibly specify additional holes to meet application needs. For example, if we take all the trajectories as an arrangement of curves and place an artificial obstacle in each face of the planar decomposition, we can ensure that all trajectories are with different T-tuples – allowing the maximum level of differentiation. Alternatively, if the application asks for trajectory classification, we can possibly only mark a subset of obstacles representing major landmarks such that the trajectories are naturally grouped to provide high-level clusters. See Fig. 4 for an example.

Notice that even the T-class of two trajectories are the same, they may still have different homotopy type. Take Fig. 5 as an example, cycle $\alpha - \beta$ and cycle $\alpha - \gamma$ are in the same T-class since they both enclose loop $l_0$. However, $\beta$ cannot be smoothly deforms to $\gamma$.

## 5 Analysis of the Algorithm

In this section, we present evaluation results of our algorithm. We first examine the pre-processing procedure of Hodge decomposition under various settings, then discuss the exactness of the discretization of harmonic 1-form. At last, we discuss the safe threshold to distinguish paths with different T-tuples. Our algorithm was tested on networks deployed on a surface with $N_h$ handles or holes. Within the given network domain, nodes of size $N_v$ are uniformly distributed and communicated with each other via adjacent edges.

Our observations are summarized as follows:

1. The convergence speed of the computation of harmonic 1-form is mainly affected by the iterations of Hodge decomposition. The scale of the network affects limitedly on convergence speed.

2. The exactness of the algorithm is independent of the randomly generated initial value.

3. With properly chosen threshold based on the exactness requirement, our trajectory classification algorithm is guaranteed to be correct.

## 5.1 Impact of Randomness

The Hodge decomposition algorithm begins with randomly generated 1-form $\omega_r$ on each edge. To show the impact of the randomness to our algorithm, we evaluate our algorithm on two different network domains. For each domain, we fix the input network ($N_h$, $N_v$), set the threshold to be $\varepsilon$, and randomly generate 100 different initial 1-forms for Hodge decomposition. Here, the precision threshold $\varepsilon$ is applied for computing the exact form $d\mathbf{f}$ and coexact form $\delta\mathbf{g}$ in Hodge decomposition process. The iteration process stops when the amount of change of $d\mathbf{f}$ and $\delta\mathbf{g}$ on each edge is smaller than $\varepsilon$. For each 1-form, the number is randomly generated within $[-1.0, 1.0]$.

In the experiment, we mainly focus on convergence speed ($\{I_{df}, I_{\delta g}\}$) and error ($\{E_{dh}, E_{\delta h}\}$) of Hodge decomposition.

- $\{I_{df}, I_{\delta g}\}$: Convergence speed: Number of iteration required on each node to compute the exact and coexact component until the given threshold.

- $\{E_{dh}, E_{\delta h}\}$: Exactness: Errors of the computed harmonic component $\mathbf{h}$. $E_{dh}$ is the error estimation of $d\mathbf{h} = 0$ and $E_{\delta h}$ is the error estimation of $\delta\mathbf{h} = 0$.

Table 1 illustrates the results of two different parameter settings, in the table, $1e-1$ means $1 \cdot 10^{-1}$. In table 1(a), we observe that $I_{df}$ has a standard deviation 5, which is 16.12% of the average value 31, $I_{\delta g}$ has a standard deviation 9, which is 14.28% the average value 63. $E_{dh}$ and $E_{\delta h}$ have standard deviation $1.2e-3$ (9.46%) and $2.5e-3$ (9.34%) around average value $1.3e-2$ and $2.7e-2$. In table 1(b), $I_{df}$ and $I_{\delta g}$ show a larger percentage of deviation, while $E_{dh}$ and $E_{\delta h}$ still show a similar percentage of deviation as in table 1(a).

In summary, the randomness while initializing $\omega_r$ could incur considerable deviations (varying from 14.28% to 33.80%) in the number of iterations $\{I_{df}, I_{\delta g}\}$, while only small fractions of deviations (less than 10%) in the harmonic 1-form error $\{E_{dh}, E_{\delta h}\}$.

**Table 1.** Performance of Hodge decomposition under the impact of randomness of input 1-forms.

(a) $N_h = 6$, $N_v = 3000$, $\varepsilon = 5e-3$, $\#test = 100$.

|  | $I_{df}$ | $I_{\delta g}$ | $E_{dh}$ | $E_{\delta h}$ |
|---|---|---|---|---|
| $min$ | 22 | 48 | 0.975e-002 | 2.024e-002 |
| $max$ | 51 | 97 | 1.471e-002 | 3.186e-002 |
| $avg$ | 31 | 63 | 1.297e-002 | 2.682e-002 |
| $std$ | 5 | 9 | 1.228e-003 | 2.507e-003 |
| $std/avg(\%)$ | 16.12 | 14.28 | 9.46 | 9.34 |

(b) $N_h = 3$, $N_v = 3000$, $\varepsilon = 5e-4$, $\#test = 100$.

|  | $I_{df}$ | $I_{\delta g}$ | $E_{dh}$ | $E_{\delta h}$ |
|---|---|---|---|---|
| $min$ | 101 | 253 | 1.151e-003 | 2.412e-003 |
| $max$ | 465 | 631 | 1.492e-003 | 3.561e-003 |
| $avg$ | 210 | 388 | 1.400e-003 | 2.921e-003 |
| $std$ | 71 | 79 | 7.385e-005 | 2.758e-004 |
| $std/avg(\%)$ | 33.80 | 20.36 | 5.27 | 9.44 |

To separate the impacts of all parameters from the randomness of $\omega_r$, for the remaining experiments we apply the average results over 100 random simulations, and the random numbers are generated from $[-1.0, 1.0]$.

### 5.2 Convergence Speed of Hodge Decomposition

The convergence speed of Hodge decomposition is measured by the number of iterations $\{I_{df}, I_{\delta g}\}$. During each iteration, one node communicates with all of its neighbor nodes to refine the 0-form **f** and 2-form **g**, this iteration process stops only when the exactness requirement of these two forms is achieved.

Fig. 6 demonstrates the convergence speed under different variables: the number of nodes $N_v$; number of holes $N_h$; and exactness parameter $\varepsilon$. We observe that $I_{df}$ and $I_{\delta g}$ are reversely proportional to $\varepsilon$, while only slightly impacted by $N_v$, and almost not impacted by $N_h$. Notice that since the computations in each iteration consist only lightweight operations(Equation 4 and Equation 5), the total computational time of all experiments remains within seconds.

### 5.3 Accuracy of Hodge Decomposition

The accuracy of Hodge decomposition is measured by $\{E_{dh}, E_{\delta h}\}$. The smaller the $\{E_{dh}, E_{\delta h}\}$ are, the more accurate the resulting harmonic form is. Fig. 7 presents how these errors vary over networks with various holes $N_h$, number of nodes $N_v$ and exactness threshold $\varepsilon$. In summary, $E_{dh}$ and $E_{\delta h}$ grow proportionally over $\varepsilon$, but almost keep constant over $N_h$ and $N_v$.

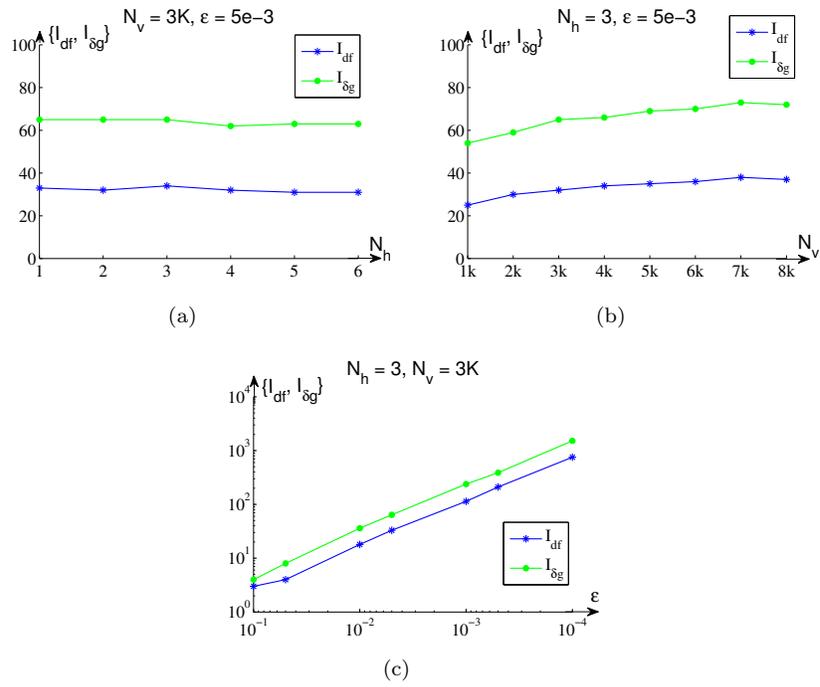

**Figure 6.** Average number of iterations $\{I_{df}, I_{\delta g}\}$ of the Hodge decomposition process v.s. number of holes $N_h$ (Fig. 6(a)), number of nodes $N_v$ (Fig. 6(b)) and threshold of exactness $\varepsilon$ (Fig. 6(c)). Here $1k$ means 1000 nodes.

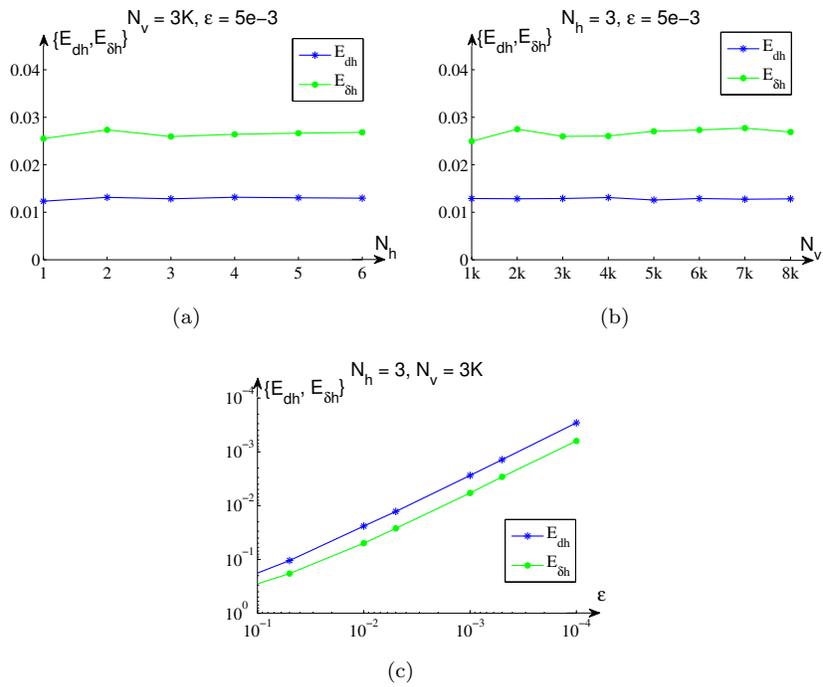

**Figure 7.** Error $E_{dh}$, $E_{\delta h}$ of the Hodge decomposition results v.s. number of holes $N_h$(Fig. 7(a)), number of nodes $N_v$(Fig. 7(b)), and exactness threshold $\varepsilon$(Fig. 7(c)). Here $1k$ means 1000 nodes.

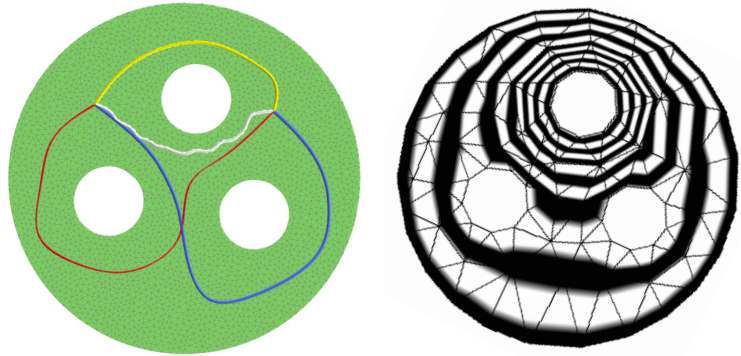

(a) 4 homology classes (5 paths each)  (b) A harmonic 1-form on a graph with low density of 85 nodes.

**Figure 8.** Correctness of path homology detection. Experiment is carried out on a 3-hole disk domain with 3000 nodes.

## 5.4 Correctness of Trajectory Classification

The trajectory classification algorithm takes a harmonic 1-form basis $\Omega = \{\omega_1, \cdots, \omega_m\}$ and a user-specified threshold $\mu$ to check if two input paths $\gamma_1$ and $\gamma_2$ are with the same T-tuple. The vector $h$ in a T-tuple $(s, t, h(\gamma))$ encodes the evaluation of $\Omega$ on $\gamma$. That is, the $k$th element of $h$ is $\langle \omega_k, \gamma \rangle$. Moreover, since

$$\langle \omega_k, \gamma_1 - \gamma_2 \rangle = \langle \omega_k, \gamma_1 \rangle - \langle \omega_k, \gamma_2 \rangle,$$

we can estimate the difference of $h$ between $\gamma_1$ and $\gamma_2$ by

$$\sigma(\gamma_1, \gamma_2) = \max_{\omega_k \in \Omega} |\langle \omega_k, \gamma_1 - \gamma_2 \rangle|.$$

We say that $\gamma_1$ and $\gamma_2$ are with the same T-tuple if and only if $\sigma(\gamma_1, \gamma_2) < \mu$. The correctness of trajectory classification heavily depends on the choice of the threshold $\mu$. If $\mu$ is too small, the paths supposed to be in the same T-class might be misclassified into different classes; If $\mu$ is too large, the paths from different T-classes might be incorrectly unified into a T-class.

To investigate a proper $\mu$, we arbitrarily choose $n = 20$ paths $\{\gamma_1, ..., \gamma_n\}$ with 4 different T-tuples on a planar domain with 3 holes and 3000 nodes shown in Fig. 8(a). Each pair of trajectories $(\gamma_i, \gamma_j)$, $1 \leq i < j \leq n$ is classified correctly when we adopt $\mu \in [1e^{-5}, 1e^{-4}]$. In practice, our trajectory classification algorithm is fairly robust against the density of nodes. Take Fig. 8(b) as an example, in the domain with only 85 sensor nodes, the experiment in Fig. 8(a) still works successfully.

# 6 Evaluation

In this section, we evaluate our trajectory classification algorithm under one model layout and one real world data. For the first data, we test our algorithm with randomly generated trajectories on the floor plan of a gallery; for the second data, we analysis the taxi trajectories collected in Shenzhen City, China, and classify them into different categories.

## 6.1 Trajectories in A Gallery

For a given museum, a gallery or an exhibition center, we are interested in the visiting patterns of visitors: preferred rooms, preferred object visiting order, and do they like to visit the room in clockwise order or counter-clockwise order? Suppose all visitors enter and leave the domain from the same entrance and exit. If we treat walls that separated the rooms as holes in a domain, the features of these visitors' patterns, such as the room visiting order, room visiting orientations, are actually the T-tuples of these trajectories. Hence by our algorithm, we are able to classify all these trajectories into different T-tuples.

We first choose the domain to be the floor plan of a museum, as shown in Fig. 9. In this domain, we uniformly distribute 3625 nodes with the average degree of 5.565; the domain (museum) is separated by obstacles (walls) into 15 rooms, entrance and exit are marked as circle and triangle, respectively.

To detect trajectories with different T-tuples, random trajectories are generated in a two level manner. We first decide room sequences to walk through, then construct the trajectories based on these room sequences. For the first level, we treat each room in the domain as a node, and connect two room nodes if they share the same door. In this way, the domain is converted to a topology graph of square nodes and dotted lines as seen in Fig. 9. The visiting room sequences are randomly chosen from the entrance to the exit. For the second level, we randomly choose a point for each room in the sequence as an intermediate node to pass through, then connect these nodes in order with shortest paths.

Fig. 10 shows different trajectories with three different T-tuples grouped by our algorithm. In this figure, sample trajectories enter the domain from the red circle and exit to the red triangle. Trajectories with the same T-tuple imply that they all bypass the obstacle with the same order and same direction. In this museum example, all trajectories can be simply represented by T-tuples, and our algorithm is able to differentiate all 32 possible T-tuples (5 obstacles with $2^5$ possibilities for simple paths).

Not limited to 2D domain, our algorithm also performs well on multi-dimensional high genus surfaces. In Fig. 11, we extend the museum to 3 levels with 2 extra ladders connecting each level. To compute the harmonic 1-form of this 3-level floor plan, we apply double covering to convert it into a closed surface with 20 handles.

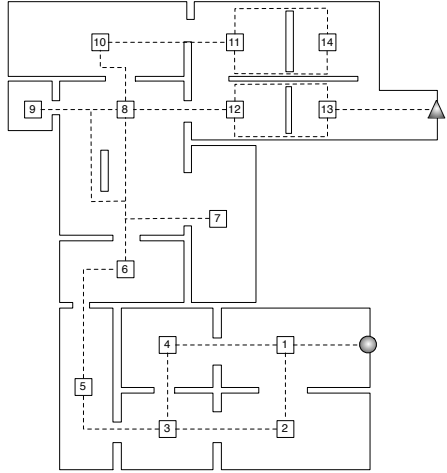

**Figure 9.** The floor plan of a museum. The entrance and exit are marked as circle and triangle, respectively. A higher level of room topology graph is represented by the square in each room and the dotted lines.

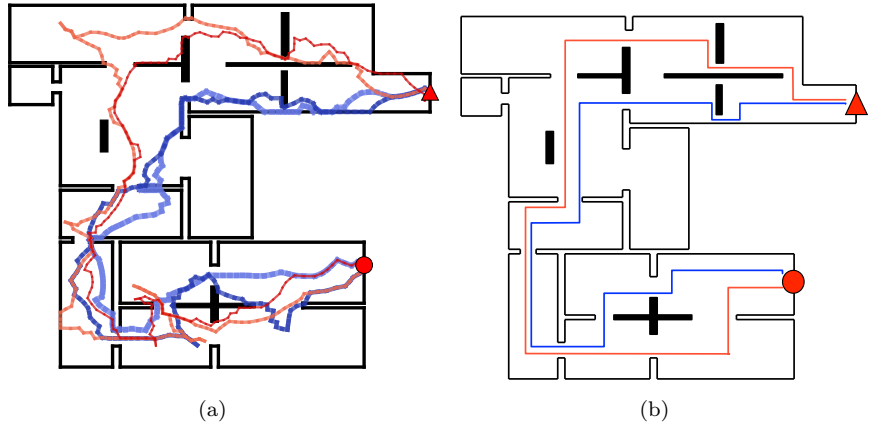

**Figure 10.** A demonstration of trajectories of 2 different T-tuples in a museum. Trajectories with the same T-tuples travel rooms with the same sequence and go around the walls in the same direction. In Fig. 10(a), trajectories with same T-tuple are labeled with same color, their simplified trajectories are displayed in Fig. 10(b).

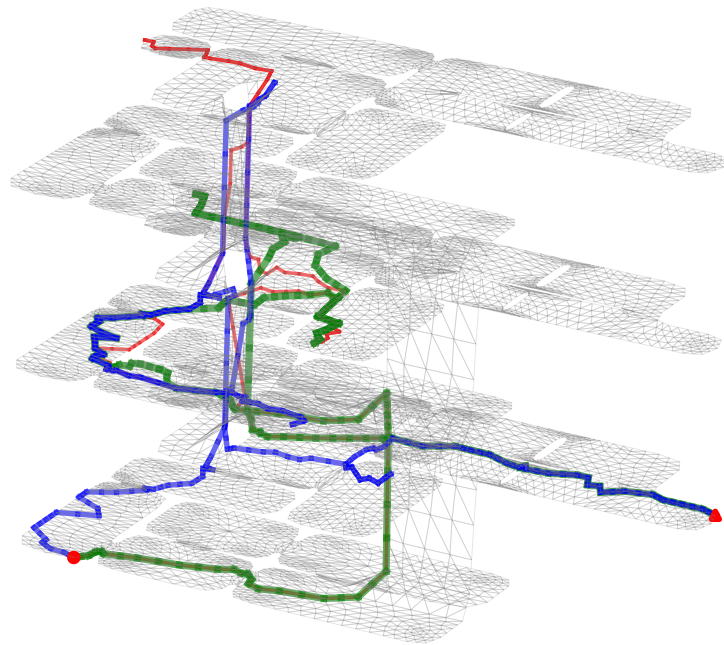

**Figure 11.** A demonstration of trajectories with three different T-tuples in a 3 floors museum. This complex surface is extended from the floor plan of Fig. 10 by connecting each level with 2 ladders, consisting 13626 sensor nodes with the average degree of 6.0.

Table 2. Taxi data in Shenzhen.

| Trajectory Data Before Processing | | | |
|---|---|---|---|
| longitude | latitude | #trajectory | sample points |
| 111.92∼ 116.76 | 21.52∼ 23.47 | 9386 | 288 |
| Trajectory Data After Processing | | | |
| longitude | latitude | #trajectory | average points per trajectory |
| 114.11∼ 114.14 | 22.54 ∼ 22.57 | 243 | 21.6 |

Table 3. Descriptive nature of T-tuples.

| #holes | #T-tuples | max. # trajectories in the same T-tuple | #trajectories in unique T-tuple |
|---|---|---|---|
| 3 | 41 | 48 | 21 |
| 5 | 105 | 26 | 69 |
| 7 | 146 | 22 | 119 |

## 6.2 Taxi Trajectories in Shenzhen

The taxi trajectory data are collected from 9386 taxis in Shenzhen, the location of each taxi is sampled for every 5 minutes for one day. The trajectory of each taxi is then represented by a line segment connecting all consecutive sample points. To simplify our data, we only choose parts of the city as the sample area, all data points outside the sample area are ignored. The sample data description is given in Table 2.

We choose two frequent visited locations as source and destination points, marked as circle and triangle in Fig. 12. We then choose trajectories that going through the source and destination points for analysis. In total, we choose 243 sample trajectories, each trajectory has 21.6 sample points in average.

In this domain, 7 areas are selected as holes in the experiment, shown in Fig. 12. The trajectories of all taxis are plotted in Fig. 13(a), with line width representing the number of taxis passing through. Notice that the trajectories of taxis have specific characteristics. Intuitively, the trajectory of a taxi consists of two types of segments: with customers and without customers. The segment of the trajectory from the location where a customer is picked up to the respective destination is likely to follow a near shortest path. The segment for which the taxi carries no customers is possibly more random and may have loops or detours. This assumption matches with what we observed from the data – the trajectory from our chosen source to the destination could deviate a lot from the shortest path.

Fig. 13(b) and Fig. 13(c) illustrate 4 example trajectories with different T-tuples as Table 4. To test if two trajectories belong to the same T-class, we simply subtract one T-tuple from the other to form a cycle, and check if this cycle can shrink to a point. In Fig. 13(b), $T_7 - T_{191} = [0, 0, 0, -1, 0, 0, 0]$, which means we connect the start point of $T_{191}$ with the end point of $T_7$, and connect the end point of $T_{191}$ with the start point of $T_7$ to form a cycle. In the cycle $T_7 - T_{191}$, the value $h_3 = -1$ means this cycle winds hole $l_3$ clockwisely for 1 loop. Notice that a cycle can wind holes for multiple loops. For example in Fig. 13(c), cycle $T_7 - T_{224} = [0, 3, 2, 1, 1, 1, 1]$ means this cycle winds hole $l_1, l_2, l_3, l_4, l_5, l_6$ counter-clockwisely for 3, 2, 1, 1, 1, 1 loops, respectively.

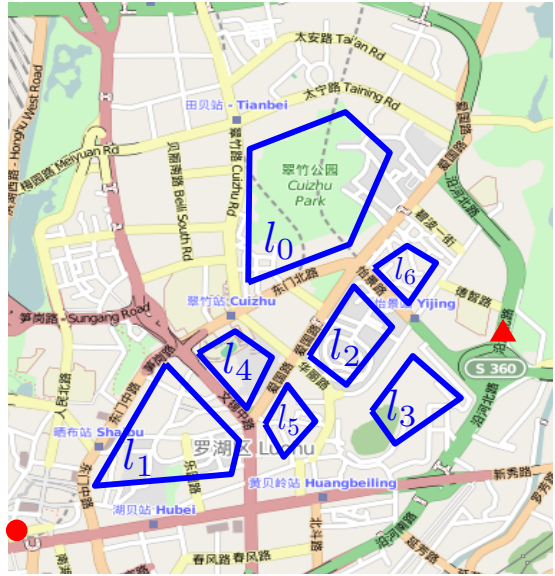

**Figure 12.** Sample area map of taxi trajectory data in Shenzhen. The holes are plotted and labeled on the real map. The source and destination are marked as circle and triangle, respectively.

The results of the number of T-tuples with multiple holes are reported in Table 3. As expected, the number of T-tuple increases quickly with the growth of the number of holes. With 7 holes, we can successfully detect 146 different T-tuples and 119 unique T-tuples. After classification, if there is a T-tuple with only one trajectory, the trajectory can be simply represented by T-tuple itself. This means that 119 out of a total of 243 trajectories can be uniquely identified with our method. This shows the descriptive power of using T-tuples to represent trajectories. With the compact signature of merely 7 numerical numbers, we can narrow down each trajectory into buckets, out of a family of totally 243 trajectories.

**Table 4.** T-tuples of trajectories in Shenzhen examples. ($h$ part only)

| Trajectory | T-tuple ($[h_0, h_1, h_2, h_3, h_4, h_5, h_6]$) |
|---|---|
| $T_7$ | [0.251484, 0.529039, 0.460121, -0.376385, 0.471286, 0.5539286, 0.352638] |
| $T_{191}$ | [0.251484, 0.529040, 0.460121, 0.623614, 0.471286, 0.553928, 0.352638] |
| $T_{196}$ | [0.251484, 0.529040, 0.460120, -0.376385, 0.471286, -0.446071, 0.352638] |
| $T_{224}$ | [0.251484, -2.470960, -1.539879, -1.376385, -0.528713, -0.446071, -0.647361] |

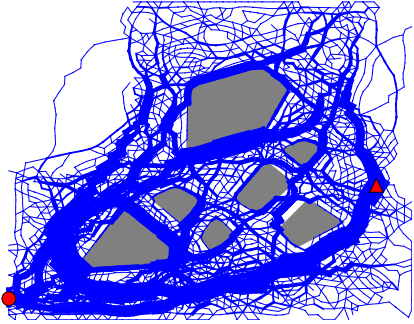

(a) Trajectory flow

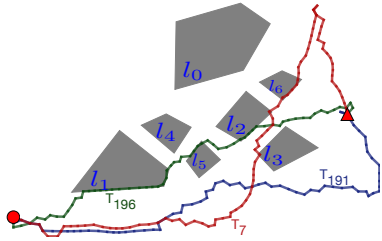 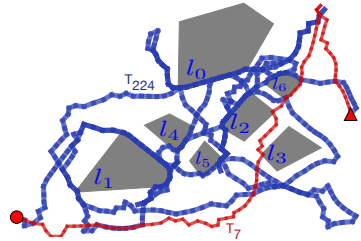

(b) 3 trajectories with different T-tuples       (c) 2 trajectories with different T-tuples

**Figure 13.** In Fig. 13(a), all trajectories we sampled are plotted as a flow where the width represents the number of taxis going through the path. In Fig. 13(b) and Fig. 13(c), 4 trajectories with different T-tuples are plotted with different color; the T-tuples of these trajectories are listed in Table 4.

# 7  Conclusion

In this paper, we use Hodge decomposition to compute a set of distributed, harmonic 1-forms on edges of a network. The harmonic 1-forms encode the topological information of the network and thus can be used for detecting and categorizing trajectories. This allows us to group trajectories into T-tuples and thus provide a meaningful way of clustering. We evaluate our algorithm with randomly generated trajectories in an art museum layout and real world taxi trajectories in Shenzhen City. Our algorithm is shown to be effective to correctly detect, analyze, and classify all given trajectories. We expect to apply our method for larger scale of trajectories analysis in near future.